\documentclass[aps,prl,twocolumn,groupedaddress]{revtex4-1}

\usepackage[dvips]{graphicx}

\begin{document}


\title{Role-similarity based comparison of directed networks.}


\author{Kathryn Cooper}

\author{Mauricio Barahona}
\affiliation{Department of Bioengineering \& Institute for Mathematical Sciences, Imperial College London, South Kensington Campus, London SW7 2AZ, United Kingdom}


\date{\today}

\begin{abstract}
The widespread relevance of complex networks is a valuable tool in the analysis of a broad range of systems.  There is a demand for tools which enable the extraction of meaningful information and allow the comparison between different systems.  We present a novel measure of similarity between nodes in different networks as a generalization of the concept of self-similarity.  A similarity matrix is assembled as the distance between feature vectors that contain the in and out paths of all lengths for each node.  Hence, nodes operating in a similar flow environment are considered similar regardless of network membership.  We demonstrate that this method has the potential to be influential in tasks such as assigning identity or function to uncharacterized nodes.  In addition an innovative application of graph partitioning to the raw results extends the concept to the comparison of networks in terms of their underlying role-structure.
\end{abstract}


\maketitle


The study of complex networks has experienced a dramatic rise in popularity in recent years.  Networks are a valuable way of representing complex data from a broad range of systems~\cite{Girvan:2002p46, Strogatz01, Newman:2003p39, Boccaletti2006175}.  Prominent examples include the World Wide Web, protein interaction networks, food webs and political blogs.  A network is a collection of nodes (individuals, web-pages, species, proteins etc) connected by edges that represent interactions (friendships, hyperlinks, predation, correlated behavior etc).  

With the escalation in computational capability and high-throughput technologies, networks can contain millions of nodes or more, and useful information must be gleaned by means of statistical properties rather than by the study of each individual.  In response to the demand for data analysis techniques, a huge variety of measures have been proposed to quantify and compare properties of these systems.   

A key challenge in this area is the development of methods to obtain simplified representations or models of complex network structure that capture the important characteristics.  Initial models with random connections do not display features common to many real world networks \cite{Amaral:2006p4086}.  For instance, many networks exhibit modular structure and numerous tools have been developed to detect tightly knit communities of significantly related nodes~\cite{Fortunato201075, Newman:2006p26}.  

Almost all network research to date has focused on simple network relations.  Edges are binary and without information on direction or value of each interaction.  For a large subset of data the direction of interaction is essential.  Examples include predation in food webs, hyperlinks in the world wide web, and systems involving causality such as metabolic networks. 

On the whole, the direct comparison between different networks has so far been restricted to general statements summarizing network statistics.  Examples of these measures include degree distribution and connectivity.  There is far more to be uncovered and here we focus on the cross-network comparison of the role played by individual nodes and groups of nodes within network structure.  Here, we will focus on the development of these methods and demonstrate the implications of such an investigation.  For one, this enables the identification of functionally equivalent individuals within different networks.  Secondly this gives rise to the development of a measure of network similarity based upon the underlying structure.


The subject of role similarity has been addressed from a number of angles and we take inspiration from a variety of sources.  The social sciences have provided the impetus for much of this research with concepts such as centrality~\cite{Katz1953, Borgatti:2005p7},  regular equivalence and block modeling~\cite{Borgatti:1993p2749}.  The research behind search engine algorithms \cite{Kleinberg:1999, Page:1999p4263, Fortunato:2006p4317}  is also highly relevant.  From a graph theoretical perspective there have been approaches by Jeh and Widom \cite{SimRank}, Blondel et al \cite{Blondel:2004p2427} and Leicht, Holme and Newman \cite{Leicht:2005}. Here we make a direct extension to our recently developed method for self-similarity calculation ~\cite{CooperBarahona1}, the purpose being to measure the functional similarity between nodes in different networks based upon network connections.

Given that the concept of role is dependent on the flow of information through a system, the inclusion of edge direction is natural.  Regardless of which network a node belongs to, a flow profile is compiled using powers of the relevant adjacency matrix. A scaling parameter tunes the relative importance of local and global information.  For instance, using only the most local edge information, sources and sinks can be identified in any network.  For two nodes to be considered similar their networks will appear similar from their perspective and as more distant information is included a more detailed  structure emerges.  Of course, a specific case is that in which the two networks are the same and the method reduces to computing self-similarity.

\textbf{Method.} 
The measure is defined as follows.  Consider two directed graphs A and B with $N_a$ and $N_b$ nodes and adjacency matrix $A_a$ and $A_b$, which are in general asymmetric.  The number of outgoing paths of length $k$ for node $i$ is given by the $i$-th coordinate of the vector 
$[A^{k}\mathbf{1}]$, where $\mathbf{1}$ is the $N \times 1$ vector of ones.  Similarly, the number of incoming paths of length $k$ for node $i$ is: 
$[{A^T}^k\mathbf{1}]_i$. Note that the case $k=1$ corresponds to the out-degree and in-degree which, from this perspective, represent the number of paths of length one originating or terminating at the node.

For each network we construct a matrix that compiles the incoming and outgoing paths of all lengths up a maximum $K$ by appending the column vectors indexed by path length and scaled by the factors $\beta^k$: 

$$
X_a=\left [  \begin{array}{c}  \mathbf{\vec{x}}_1  \\ \vdots \\ \mathbf{\vec{x}}_N  \end{array} \right]  \equiv
\left [\begin{array}{cccc|cccc} &&&&&&&\\ [0.4em]
\mathbf{ \vec{v}_1} & \mathbf{ \vec{v}_2} & \ldots & \mathbf{\vec{v}_K} & \mathbf{\vec{w}_1} & \mathbf{\vec{w}_2} & \ldots & \mathbf{\vec{w}_K}\\ [0.4em]
&&&&&&&
 \end{array}  \right] 
$$

Where the column vectors $\mathbf{\vec{v}_k}=(\beta A)^k \mathbf{\vec{1}}$ and $\mathbf{\vec{w}_k}=(\beta A^T)^k \mathbf{\vec{1}}$.  $\beta=\alpha/\lambda_{1a}$, with $\lambda_{1a}$ the largest eigenvalue of the adjacency matrix and $0 \leq \alpha \leq 1$. The parameter $\alpha$ is a scaling factor that allows us to tune the weight of the local environment (short paths) relative to the global network structure (long paths). The presence of the factors $\beta^k$ ensures the convergence of the sequence of the columns due to the asymptotic limit $\lim_{k \to \infty} \frac{||A^{k+1}||}{||A^k||} \rightarrow \lambda_1$. $K$ is defined as the point at which the columns have converged \cite{Katz1953}.

Each row vector of $X_a$ and $X_b$ contains the flow profile of a node in terms of the scaled number of incoming and outgoing paths of all lengths starting and ending at that node.  The similarity between two nodes regardless of the network to which they belong can be quantified by the distance between the vectors $\mathbf{x_a}_i$ and $\mathbf{x_b}_i$.   A simple choice of metric is the cosine distance, which leads to the (generally) rectangular similarity matrix $Y$ defined by:
\begin{equation}
Y_{ij}= \frac{\mathbf{x_a}_i  \mathbf{x_b}^T_j}{||\mathbf{x_a}_i|| \, ||\mathbf{x_b}_j||},
\end{equation}

where element $Y_{ij}$ provides a normalized measure of the closeness of the flow profiles of node $i$ in network A and node $j$ from network B.   Naturally, if one compares a network with itself then the resulting matrix will be square with a diagonal consisting of ones.

The above similarity matrix can also be calculated via an iterative algorithm based upon a definition of node similarity, making the method directly comparable to previously proposed algorithms and offering additional insight into the concept.

\begin{figure}[htp]
\centering
\includegraphics[]{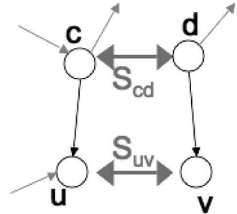}
\caption{The intuitive concept behind the iterations is that the similarity between nodes u and v is determined by their immediate connections and by the similarity between their neighbours.\label{fig:iterativefig1}}
\end{figure}

A pair of similar nodes will be connected in the same way to other pairs of similar nodes.  Referring to figure \ref{fig:iterativefig1},  if c and d are similar, and u is connected to c, and v is connected to d in the same way, then we can say that nodes u and v also have some similarity.  The similarity between two nodes is determined by both the nature of their immediate connections ($AJA^T$) and also the similarity of their neighbours ($AY_nA^T$) .    The primary source of similarity is the immediate neighbourhood and the relative importance of the local connections in comparison to the external information passed from a node's neighbours is determined by tuning the parameter $\alpha$.

The result is the following iterative procedure for calculating a matrix of similarity scores, $Y$, with $Y_0=$ the matrix of zeros$(N_a,N_b)$.

Considering outgoing connection $Y(OUT)$ is defined as the convergent term of the iteration:

\begin{equation}
Y_{n+1}=A_a(J+\frac{\alpha^2}{{\lambda_a}_1{\lambda_b}_1}Y_n)A_b^T
\end{equation}

where $J$ is the matrix of ones and $Y_0$ is the matrix of zeros. Likewise for incoming connections $Y(IN)$ is defined as the convergent term of the iteration:

\begin{equation}
Y_{n+1}=A_a^T(J+\frac{\alpha^2}{{\lambda_a}_1{\lambda_b}_1}Y_n)A_b
\end{equation}

The final Similarity matrix is the sum of the final convergent terms of $Y(OUT) + Y(IN)$ after a normalisation procedure.   Convergence is guaranteed by the presence of the $\frac{\alpha^2}{\lambda_a\lambda_b}$ term. 

The normalisation procedure for rectangular matrices is to divide each entry of the AB-similarity matrix, $Y_{ij}(A,B)$ by the square root of the diagonal entries of the respective self-similarity matrices,  $\sqrt{Y_{ii}(A,A)}$ and $\sqrt{Y_{jj}(B,B)}$.   Thus each entry is normalised by the value that would be produced if the node were compared to itself.

This algorithmic formulation allows for simplified updated computations in a format equivalent, yet functionally distinct, to other methods~\cite{Blondel:2004p2427}.  The method described above neatly reduces to the self-similarity measure \cite{CooperBarahona1} and unlike previous methods does not suffer from issues arising from odd and even iterations \cite{Leicht:2005}.  Convergence of the algorithm is guaranteed naturally and no prior assumptions regarding any similarity between nodes is required.  Structurally equivalent nodes will have similarity of one, regardless of the ordering of rows and columns.

\begin{figure}[htbp]
\centering
\includegraphics[]{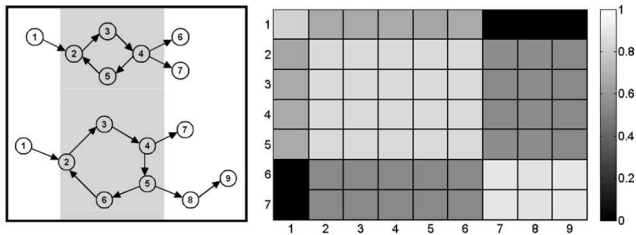}
\caption{Small example networks and the resulting similarity matrix from $\alpha=0.99$ represented in gray scale.
\label{fig:smallEx} }
\end{figure}

As an illustration, we explore a comparison between two small example networks.  We construct two small directed networks as shown in figure \ref{fig:smallEx}.  The resulting rectangular similarity matrix is also shown in gray-scale, indicating groups of nodes from both networks which appear to play a similar role in their respective environments.

The choice of the parameter $\alpha$ produces the most interesting results at values close to 1.  At lower values we trivially compare only the degree of each node. Care must be taken at the high end of the spectrum as values too close to 1 tend to be computationally prohibitive.  As a rule of thumb, $0.95 \leq \alpha \leq 0.99$ produce interesting and stable results.

\textbf{Results.}  
Several alternatives exist to process the information contained in the resulting similarity matrix for larger, real-world networks.  We choose two complimentary lines of attack: the extraction of quantitative observations directly from the similarity matrix, and further processing in the form of partitioning the rectangular matrix.

\textbf{Direct use of matrix entries.}  
The first approach is to simply use the raw information.  This in itself is very informative.  For instance we can identify the node(s) in network $B$ that most closely match a specific node from network A. This can be applied to assign a putative function or classification to nodes in an uncharacterised network when compared to a network in which the node identity is known.

The potential application of this method is demonstrated on a small selection of metabolic networks.   The database compiled from KEGG ~\cite{kanehisa2000} by Ma and Zeng \cite{Ma2003} includes 80 fully sequenced organisms in an extensive and carefully revised bioreaction database. Crucially, these networks have been represented as directed graphs by taking into account the reversibility of reactions.  This data has been found to exhibit core-periphery type structure by various methods \cite{daSilva:2008p4081, Holme:2005p4080, CooperBarahona1}

The identity and characteristics of each node (metabolite) is already well studied.  We take a small selection of these networks: \textit{Bacillus subtilis} (bsu), \textit{Listeria monocytogenes} (lmo),  \textit{Escherichia Coli} (eco), \textit{Homo sapien} (hsa), \textit{Mus Musculus} (mmu),  \textit{Pseudomonas aeruginosa} (pae), \textit{Pyrobaculum aerophilum} (pai) and \textit{Sulfolobus solfataricus} (sso).

\begin{figure}[htbp]
\centering
\includegraphics[]{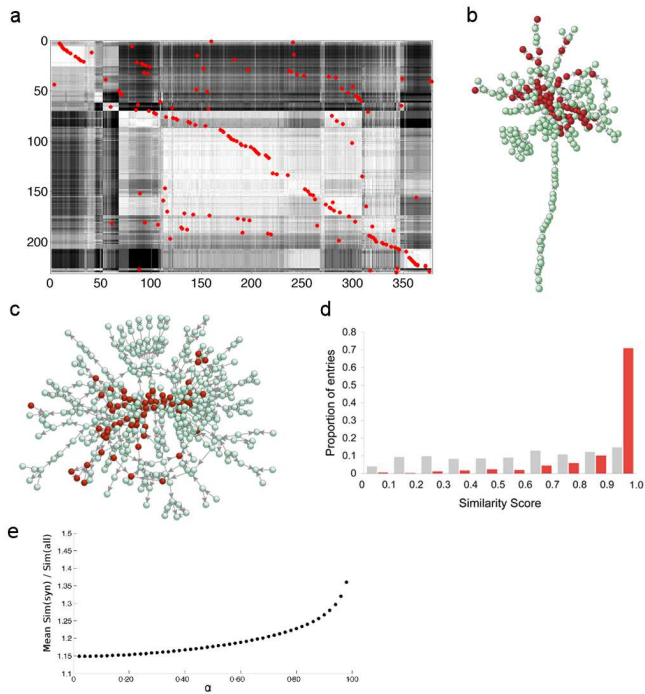}
\caption{(a) Similarity matrix between metabolic networks lmo (b) and pai (c), rearranged to approximately maximize the diagonal, with synonymous node pairs highlighted in red.  The distribution of matrix entries over all network pairs is shown in histrogram (d).  In all four diagrams gray scale represents all nodes and red represents the synonymous metabolite pairs. (e) The mean similarity between synonymous nodes with respect to the mean matrix entry, averaged over all network pairs, increases with $\alpha$.  \label{fig:idmetabolites}}
\end{figure}

As the metabolic identity of each node is known, we can compare the role played by synonymous nodes (i.e. the same metabolite) across multiple networks.  Figure \ref{fig:idmetabolites} illustrates the concept for the metabolic networks \textit{lmo} (\ref{fig:idmetabolites}b) and \textit{pai} (\ref{fig:idmetabolites}c).  The metabolites common to all 8 networks are highlighted in red.  These two networks share 195 common metabolites.  With the rows and columns of the similarity matrix reordered to maximize the ``diagonal'' and the synonymous metabolites highlighted in red figure \ref{fig:idmetabolites}a highlights two interesting observations.  Firstly, on the whole, metabolites tend to have a high similarity score with themselves (they are close to the diagonal).  Secondly, they are spread throughout the network, with the complete range of roles represented. 

Over all network pairs in our set, the distribution of similarity scores for synonymous nodes (red) compared to the distribution of all entries in the matrix (gray) is shown in figure~\ref{fig:idmetabolites}d.  Synonymous nodes tend to have high similarity scores, implying that most metabolites play a similar role in each network. The average score for synonymous metabolites across networks is significantly higher ($0.880\pm 0.093$) than the overall average ($0.628\pm 0.047$).

Furthermore, on scanning through the complete range of the scaling parameter $\alpha$, we find that the mean similarity score of synonymous nodes with respect to the overall matrix mean increases as $\alpha$ tends towards its maximum (figure \ref{fig:idmetabolites}e).  Although the local environment (largely the degree) of each node is informative, there is much to be gained from the inclusion of a global perspective.

From this part of the analysis, the role in a metabolic sense appears to be correlated with our definition in terms of network flow.  This has clear and promising implications for the use of these techniques in function or identity prediction.

%

\textbf{Partitioning the similarity matrix.}  
An additional level of processing can extract further information from the matrix by grouping nodes according to their similarity score.  The method involves the simultaneous clustering (co-clustering) of nodes in network A and nodes in network B ~\cite{Dhillon:2001p4248}.  It is a simple extension of the popular spectral partitioning method, the normalised cut to a bipartite graph by using singular value decomposition. \cite{Shi:2000p825}

The result of partitioning the rectangular similarity matrix is two vectors describing the grouping of nodes in each network:  a partition of network A as classified by comparison with network B and vice versa.  It is possible that structure may be revealed within a network via comparison with another that is not obvious with self-similarity analysis.

This leads towards a measure of similarity between networks.  If we find the partition vectors of both networks as obtained from the rectangular similarity matrix $Y(A,B)$ are comparable to the partition vectors obtained from the respective self-similarity matrices ($Y(A,A)$ and $Y(B,B)$) then we could consider the underlying role organization of the two networks to be similar.  As such we propose a method of comparison between two networks based upon the partition vectors obtained.  We utilize a well studied quantity known as mutual information \cite{mutualinfo, mutualinfo2} which measures how well two partition vectors describe the same data.

The normalised Mutual Information between partitions A and B is given by:
\begin{equation}
MI(A,B)=\frac{-2\sum_{i=1}^{c_A}\sum_{j=1}^{c_B}N_{ij}log\big(\frac{N_{ij}N}{N_{i.}N_{.j}}\big)}{\sum_{i=1}^{c_A}N_{i.}log(\frac{N_{i.}}{N})+\sum_{j=1}^{c_B}N_{.j}log(\frac{N_{.j}}{N})}
\end{equation}

Where the matrix $N$ is defined as a confusion matrix between partitions A and B.  Rows of $N$ correspond to clusters in partition A and columns to clusters in B.  $N_{ij}$ is the number of objects in both cluster $i$ of partition A and cluster $j$ of partition B.  $N_{i.}$ and $N_{.j}$ are row and column sums respectively. $c_A$ and $c_B$ are the number of cluster in A and B.  In the case where A and B are identical, $MI=1$.  If A and B are independent then $MI(A,B)$ tends towards its minimum, zero.

So if the MI score between the two partitions of network A (one found by clustering $Y(A,A)$ and the other found by clustering $Y(A,B)$) and the equivalent MI score for network B are both high then one could conclude that the two networks have a similar underlying role structure.

When performing partitioning, the number of clusters we partition into can be somewhat arbitrary, method dependent or user dependent.    In fact by pre-defining the number of clusters the best match between two networks may be missed. Therefore we suggest that some flexibility in the number of partitions will give us the best idea of how similar two matrices are in terms of their partitions.  Although we apply spectral clustering it is feasible to use a number of different methods.

For each similarity matrix an ensemble of partitions are produced using spectral clustering into $c=2,3\ldots c_{max}$ groups. $c_{max}$ is defined by the user and will be dependent upon the size of the graph.  The value should be significantly smaller than the number of nodes in the network.  The MI score between each partition pair in the two ensembles is calculated and the maximum MI score is taken to be the similarity between the two matrices with respect to all possible partitions of the network induced by the matrices.

\begin{figure}[htbp]
\centering
\includegraphics[]{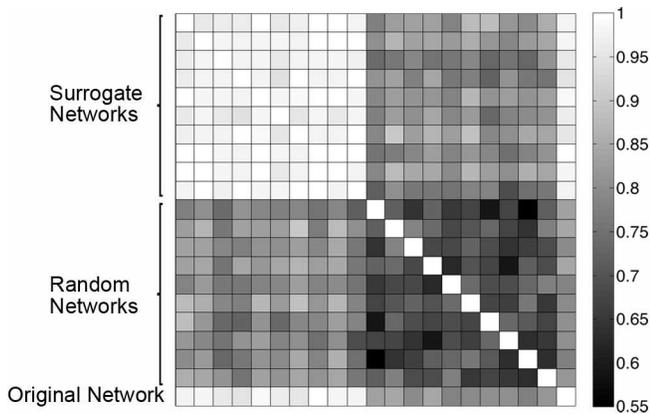}
\caption{Gray-scale representation of network similarity for world trade data and surrogate networks. The original world trade network is reduced to a three node flow representation (figure \ref{fig:cartoons}(i)f).  Surrogate networks with an identical number of nodes and edges are constructed from this model, preserving the relative flow between groups.  On comparison with the original network, with respect to random networks the surrogates show high similarity to each other and the original while the random networks perform poorly. \label{fig:WT}}
\end{figure}

\begin{figure}[htbp]
\centering
\includegraphics[]{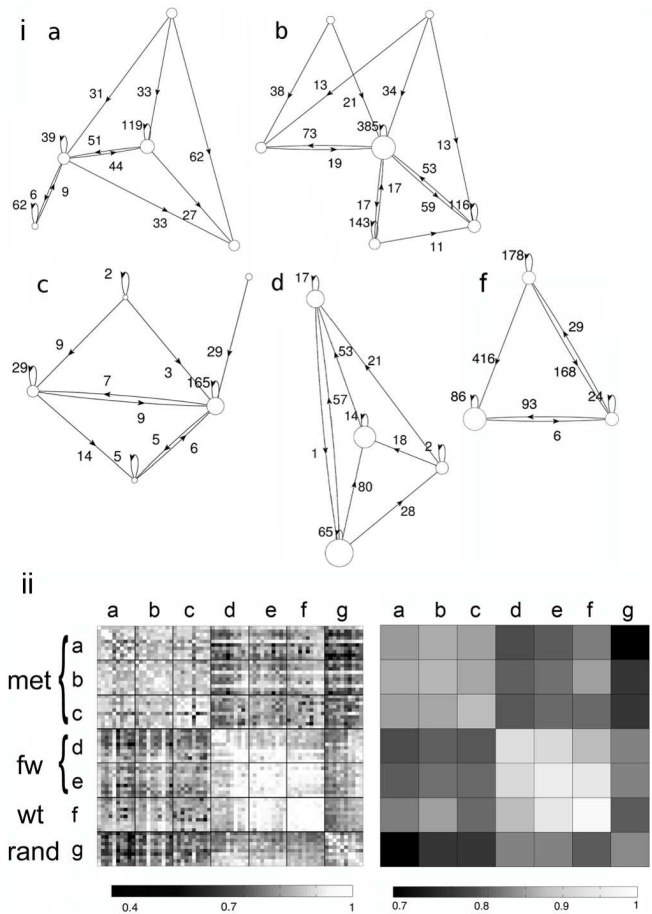}
\caption{(i)Reduced models of networks are compared by partitioning of the similarity matrices.  Metabolic network models: (a) mmu, (b) eco, (c) pai;  Foodweb models: (d) St Marks, (e) niche model (not illustrated); (f) world trade data and (g) random directed networks (not illustrated).  The representations have had edges with very low weight removed for clarity. (ii)  The resulting matrix of maximum mutual information scores is displayed in gray scale along with a block-average simplification.\label{fig:cartoons}}
\end{figure}

We have observed from the self-similarity analysis of a number of data types \cite{CooperBarahona1} that many networks can be described using a coarse-grained reduced representation by partitioning the self-similarity matrix.  We construct an ensemble of directed networks for which the proportional flow structure between groups and the group sizes is identical to the original network. All other aspects of the networks are random and in this case we keep the total number of edges the same. For example the world trade network \cite{deNooy:2005p3733, Smith:1991p3591} previously analyzed in \cite{CooperBarahona1} divides into three groups (figure \ref{fig:cartoons}f).  As a test of this reduced model, figure \ref{fig:WT} demonstrates that these surrogate networks have significantly higher similarity with each other and with the original network than completely random networks with the same number of nodes and edges. Thus, using the reduced flow representation we can create surrogates with a similar global role-structure to the original network.

To illustrate the power of this method in distinguishing between networks with different role-structure we construct an ensemble of example surrogate networks of the same size (100 nodes) although networks of different sizes can be compared. To construct with a structure akin to that of metabolic networks we use the networks of eco, pai and mmu as templates (figures ~\ref{fig:cartoons}(i)a-c).  Foodweb structure is well studied and various methods for construction of network models exist in the relevant literature.  We follow the widely accepted ``niche'' model \cite{MartinezNiche2000} and construct 10 networks.  In addition we use the reduced representation of the St Marks ecological network \cite{Baird:1998p3522} as model to construct surrogates (figure ~\ref{fig:cartoons}(i)d).  We also include surrogates from the world trade model network (\ref{fig:cartoons}(i)f). For completeness we include a set of random directed networks.

Ten of each network type are constructed as defined by the respective models.  The result of our calculations is a $70\times70$ MI matrix describing the similarity between each network pair as defined by our measure.  This matrix is illustrated in figure \ref{fig:cartoons}(ii), together with an average of each block.  There is a high degree of similarity between the model of St Marks foodweb, the niche foodweb model and the world trade model.  The models of metabolic networks are clearly more involved, however display significantly higher similarity to each other than to the other network types.  As one would expect, completely random directed networks should display no discernible structure and are dissimilar to all the network models, including other random networks.

\textbf{Conclusion.}
We have presented a generalization of the concept of node similarity to considering nodes in different networks. The environment of each node is compared in terms of directional flow.  The most useful results are those in which the most global information is included.

The raw results contained in the similarity matrices applied to metabolic networks show a promising elevation in similarity score values between nodes that are known to represent the same metabolite in both networks.  This is a good indication of the correlation between the similarity score based upon flow structure and the biological functional similarity.

An important extension developed in this work is a method by which to compare the underlying structure of networks by performing graph partitioning on the similarity matrices.  We illustrate this method by application to a selection of network models.  It is clearly shown that the networks constructed under the same regime display a high level of similarity in their underlying structure as analyzed by this method.

%

\end{document}